\documentclass[pra,%
 reprint,
 amsmath,amssymb,
 aps,
]{revtex4-2}

\usepackage{graphicx}
\usepackage{xcolor}

\usepackage{amsmath}
\usepackage{mathtools}

\usepackage{dcolumn}
\usepackage{bm}
\usepackage{float}
\usepackage{physics}

\begin{document}

\preprint{APS/123-QED}

\title{Defect-driven incoherent skin localization}

\author{Emmanouil T. Kokkinakis$^{1,2}$}

\author{Konstantinos G. Makris$^{1,2}$}%

\author{Eleftherios N. Economou$^{1,2}$}
\affiliation{$^1$
  Department of Physics, University of Crete, 70013 Heraklion, Greece
}%
\affiliation{$^2$
 Institute of Electronic Structure and Laser (IESL), FORTH, 71110 Heraklion, Greece
}%
\date{\today}

\begin{abstract}
The process of dephasing during wave evolution has traditionally been viewed as an obstacle to localization—leading to diffusion even in strongly disordered Hermitian lattices.
In contrast, here we demonstrate how the interplay of dephasing with \emph{non-Hermitian} defects can be harnessed to engineer wave localization.
Specifically, we identify a novel dynamical localization phenomenon characterized by  wavefunction accumulation at the lattice's boundary due solely to dephasing, despite globally reciprocal couplings. Furthermore, we study the incoherent skin effect arising from coupling asymmetry, and investigate the  interplay between these antagonistic localization mechanisms. By reframing dephasing from a hindrance into a tool, this study overturns established paradigms of wave localization and paves the way for novel approaches to controlling localization phenomena in non-Hermitian physics.\end{abstract}

\maketitle

\textit{Introduction.}---
Non-Hermitian physics \cite{ElGanainy2018} has garnered significant attention as a robust framework for describing open systems. Following the first observation of parity–time ($\mathcal{PT}$) symmetry in optics \cite{Ruter2010,Makris2008,ElGanainy2007,Musslimani2008,Guo2009,Regensburger2012,Hodaei2014,Konotop2016,Feng2017,Ozdemir2019}, photonic lattices became the leading platform for engineering non‐Hermitian Hamiltonians via tailored gain and loss. Beyond complex potentials, non‐Hermiticity can also arise from asymmetric couplings, as in the Hatano–Nelson (HN) model \cite{Hatano1996,Hatano1997}. The properties of the HN lattice exhibit pronounced sensitivity to boundary conditions; under open boundary conditions (OBCs), the system possesses a real spectrum and exhibits the non-Hermitian skin effect (NHSE) \cite{Zhang2022review, Lee2016, MartinezAlvarez2018, Gong2018, Yao2018}, i.e., its eigenstates are exponentially localized at one lattice's boundary. In such lattices, wave propagation is unidirectional toward one edge, a phenomenon often termed the \emph{dynamical} skin effect \cite{Longhi2015b, Longhi2015, Longhi2016, Longhi2017a, Longhi2017b, Longhi2022, Li2022, Komis2023, Kokkinakis2024}.

Recently, a plethora of studies have explored the NHSE beyond the HN model \cite{Yi2020, Li2020, Guo2021, Jiang2023, Huang2024, Lei2024, Samanta2025, Ganguly2025, Molignini2023, Yokomizo2021, Liu2021, Ammari2025, FuZhang2023}, including many-body \cite{song2019, kim2024, Mattiello2025, Suthar2025}, higher-dimensional and higher-order \cite{Lee2019b, Kawabata2020, zhang2021, Zhang2022, Kokkinakis2025}, and Floquet-driven \cite{Li2023, Ke2023, Apostolidis2025} skin effects, with emphasis on their topological properties \cite{yao2018, Lee2019, okuma2020}. Experimentally, skin effects have been realized in optics using synthetic-dimension mesh lattices \cite{Weidemann2020}, laser arrays \cite{Gao2023}, and ring resonators \cite{Liu2022}, which effectively simulate asymmetric couplings. Beyond optics, NHSEs have also been demonstrated in acoustical \cite{Zhang2021a, Zhang2021b}, mechanical \cite{li2024}, and cold-atom platforms \cite{Zhao2025}.

In a seemingly unrelated context, dephasing—i.e., the randomization of a wavefunction’s phase, effectively modeling interactions between a system and its environment—is known to induce diffusion even in strongly disordered conservative systems, thereby suppressing Anderson localization \cite{logan_1987, evensky_1990, flores_1999, yamada_1999, gurvitz_2000, kosik_2006, schreiber_2011, moix_2013, znidaric_2013, kamiya_2015, krapivsky_2014}. Remarkably, recent studies have shown that in complex-disordered non-Hermitian systems, dephasing can, rather than causing delocalization, enhance eigenmode localization \cite{longhi_2024_prl, longhi_2024_ol, Kokkinakis2024b}.

In this \emph{Letter}, we demonstrate that dynamical skin effects can be induced in non-Hermitian lattices with lossy defects via fast dephasing. We show that, despite reciprocal couplings, dephasing leads to asymmetric propagation through directional jumps between distant lattice regions, ultimately resulting in wave accumulation at one boundary. We further investigate the incoherent NHSE induced by coupling asymmetry in the HN model \cite{longhi_2024_lsa}, and analyze the interplay between these two distinct skin-effect mechanisms. Our results provide a direct route for experimentally observing dynamical {dephasing-induced} skin effects and exploring their antagonism using readily accessible photonic mesh-lattice platforms.

\textit{Model and effects of dephasing.}--- Our study begins with a one-dimensional symmetric lattice of \(N\) evanescently coupled waveguides, indexed by \(n \in \{1,2,\dots,N\}\), and described by the Hamiltonian:
\begin{equation} 
\label{hamiltonian}
H=\sum_{n=1}^{N}\epsilon_{n}\ket{n}\bra{n}+\sum_{n=1}^{N-1}\left(\ket{n+1}\bra{n}+\ket{n}\bra{n+1}\right),
\end{equation}
where we impose OBCs, \( \ket{0}=\ket{N+1} \equiv 0 \) and \(\epsilon_{n}\) denotes the on-site potential at site \(n\). Within the coupled-mode theory framework, the coherent evolution of \(\ket{\psi} \equiv \sum_{n=1}^{N} \psi_{n}\ket{n}\) is governed by the equation
$H\ket{\psi} = -i\frac{d\ket{\psi}}{dz},$
where \(\psi_n\) represents the complex amplitude of the electric field's envelope at site \(n\).  
\begin{figure}
    \centering
    \includegraphics[width=0.9\columnwidth]{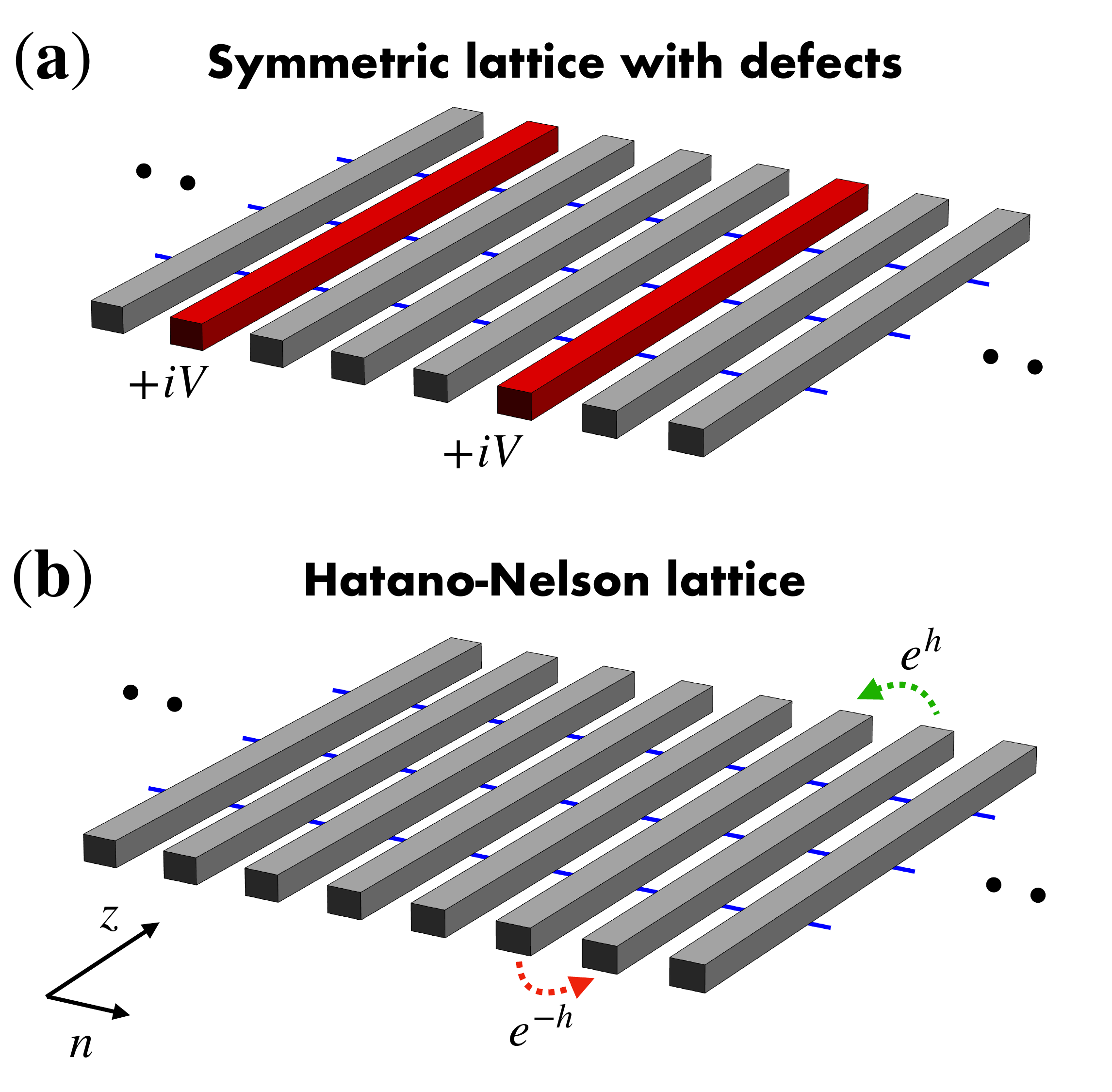}
    \caption{Schematic representation of the  waveguide lattice models examined in this work. (a) Lattice with symmetric couplings between neighboring sites, including a few on-site lossy defects. (b) Hatano–Nelson lattice with asymmetric couplings between neighboring sites. In both configurations, blue lines along the propagation direction \(z\) indicate the periodic action of \emph{dephasing}.}
\end{figure}

To incorporate dephasing into the system’s dynamics, we modify the previous equation by adding a perturbation term \(\Phi(z)\),
\begin{equation}
  H\ket{\psi} + \Phi(z)\ket{\psi} = -i\,\frac{d\ket{\psi}}{dz},
\end{equation}
which randomizes the wavefunction’s phase at each lattice site at propagation intervals \(z_{a} = a\,l\) (\(a \in \mathbb{N}\)), where \(l\) is the dephasing period. Specifically,
$\Phi(z) = \sum_{a} \Phi^{(a)}\,\delta(z - a\,l)$
with \(\Phi^{(a)} = \sum_{n} \theta_{n}^{(a)}\,\ket{n}\bra{n}\)
a diagonal matrix of uncorrelated random phases \(\theta_{n}^{(a)} \in [0,2\pi]\).
Under incoherent conditions, the evolution of the probability density \(P_{n}(z) \equiv \overline{|\psi_{n}(z)|^2}\), averaged over multiple realizations of the uncorrelated phases, is described by
\begin{equation}
\label{prop}
P_{n}(z = al) = \sum_{m=1}^{N} (\mathcal{S}^{a})_{nm} P_{n}(z=0),
\end{equation}
with incoherent propagator elements given by \(\mathcal{S}_{nm}(l) = |U_{nm}|^2\), where \(U \equiv e^{iHl}\). Since \(H\) is symmetric, \(\mathcal{S}\) is Hermitian [see Supplementary Material (SM) \cite{SM}], and its eigenvalue equation reads \(\mathcal{S}\ket{v_{j}} = \beta_{j}\ket{v_{j}}\),
with eigenvalues ordered as \(\beta_{1} \geq \beta_{2} \geq \dots \geq \beta_{N}\).

From Eq.~(\ref{prop}), we have
\begin{equation}
\label{expansion}
\ket{P(z=al)} = \sum_{j=1}^{N} d_{j,0} \beta_{j}^{a} \ket{v_{j}},
\end{equation}
where $\ket{P(z)} \equiv \sum_{n=1}^{N} P_{n}(z) \ket{n}$ and $d_{j,0} \equiv \bra{v_{j}}P(0)\rangle$. 

Thus, the projections $d_{j}(z) \equiv \bra{v_{j}}P(z)\rangle$ evolve as $|d_{j}(z)|=|d_{j,0}|\beta_{j}^{z/l}$. {A sudden \textit{jump} occurs during the dynamics when the projection \( |d_k| \) of an eigenstate \( \ket{v_{k}} \) surpasses the projection \( |d_r| \) of the previously dominant eigenstate \( \ket{v_{r}} \), under the condition $\beta_k < \beta_r$ \cite{Kokkinakis2024b, Leventis2022, Tzortzakakis2021, Weidemann2021}.} 
Consequently, after a critical distance $z_{\text{cr}}$, the eigenmode $\ket{v_{1}}$ associated with the largest eigenvalue $\beta_{1}$ dominates, yielding $P_{f}\equiv P(z>z_{\text{cr}})\propto \ket{v_{1}}$. Although \(z_{\text{cr}}\) depends on the initial excitation, the asymptotic distribution \(P_f\) is uniquely determined by the Hamiltonian \(H\) and dephasing period \(l\), irrespective of the initial condition.

For fast dephasing, i.e., \( l \ll 1 \), we can approximate \(\mathcal{S}_{kj} = U_{kj} U_{kj}^{*}\) up to $\mathcal{O}(l^2)$ as
\begin{equation}
\label{taylor}
{\small
\mathcal{S}_{kj} \approx
\begin{cases}
1 + i\Bigl( H_{kk} - H_{kk}^{*} \Bigr) l \\
\quad + \left( |H_{kk}|^2 - \frac{1}{2}\Bigl[ (H^2)_{kk} + (H^2)^*_{kk} \Bigr] \right) l^2, & k = j, \\[1ex]
|H_{kj}|^2\, l^2, & k = j\pm1.
\end{cases}
}
\end{equation}

From these relations, it first follows (see SM \cite{SM}) that for a periodic lattice ($\epsilon_{n}=0$ for all $n$), $\mathcal{S}$ is tridiagonal with diagonal terms $\mathcal{S}_{kk}=1-2l^2+(\delta_{k,1}+\delta_{k,N})l^2$ and couplings $\mathcal{S}_{k,k\pm1}=l^2$. Its eigenspectrum is given by
\begin{equation}
\label{eigenvalues}
\begin{split}
\beta_j &= 1 - 2l^2 + 2l^2 \cos\!\left[\frac{(j-1)\pi}{N}\right],\\
\ket{v_j} &\propto \sum_{n=1}^{N}
  \cos\!\left[\frac{(j-1)\pi}{N}\,\left(n-\tfrac12\right)\right]\;\ket{n}.
\end{split}
\end{equation}
with $j=1,\dots,N$. 

Therefore, the largest eigenvalue is \(\beta_{\text{1}} = 1\) with eigenstate \(\ket{v_{\text{1}}} \propto (1,1,\dots,1)^{T}\), while all other eigenvalues satisfy \(\beta_j \in (1-4l^2,1) \). As follows from Eq. \ref{taylor}, introducing real on-site deviations from periodicity (defects or disorder) leaves \(\mathcal{S}\) unchanged up to $\mathcal{O}(l^2)$ \cite{SM}. Anderson localization is therefore fragile under dephasing; even strongly disordered lattices yield uniform asymptotic probability distribution \(P_f \propto (1,1,\dots,1)^{T}\).

The dynamical behavior is significantly altered when \emph{imaginary} deviations from periodicity are introduced. Generally, in this work we will consider a lattice having a finite number \(r \ge 1\) of imaginary defects 
at sites \(q_j\) [as illustrated in Fig.~1(a)], i.e.,
$\epsilon_n = iV \sum_{j=1}^{r} \delta_{n,q_j}$ thereby being non-Hermitian (\(H \neq H^\dagger\)). 

\textit{Single non-Hermitian defect.}--- For a single imaginary defect at site \(q\), the relative difference between the defect's diagonal element and the other (bulk) diagonal elements, normalized by the coupling terms, is 
\begin{equation}
\Delta \equiv \frac{|\mathcal{S}_{qq} - \mathcal{S}_{kk}|}{l^2} = \left|2V^2 - \frac{2V}{l}\right|,
\end{equation} 
which takes huge values for \(l \ll 1\) . Consequently, the defect eigenvalue, up to $\mathcal{O}(l^2)$, is given by
\begin{equation}
\label{defect_eigenvalue}
\beta_{\text{def}}\approx \mathcal{S}_{qq} \approx 1 - 2Vl + (2V^2 - 2)l^2,
\end{equation}
and satisfies \(\beta_{\text{def}} > 1\) for gain (\(V < 0\)) and \(\beta_{\text{def}} < 1\) for loss (\(V > 0\)), 
while it corresponds to an almost single-site localized eigenvector, \(\ket{v_{\text{def}}} \approx \ket{q}\).

For a gain defect (\(V < 0\)), the defect-induced eigenmode dominates the dynamics (\(\beta_{\text{1}} = \beta_{\text{def}}\)), leading to a strongly localized asymptotic distribution \(P_f \propto \ket{q}\) after a critical distance \(z_{\text{cr}}\), which depends on the initial condition \(\psi_n(z = 0)\) \cite{SM}. 

In contrast, a lossy defect (\(V > 0\)) leads to richer dynamics, effectively dividing the lattice into two segments: sites \(1\) to \(q - 1\) and \(q + 1\) to \(N\), while the defect acts like a \textit{hard-wall}. As derived analytically in the SM \cite{SM}, the long-term probability distribution \(P_f \propto \ket{v_1}\) predominantly occupies the largest segment \(L \equiv \max(q - 1,\,N - q)\), and vanishes elsewhere, \textit{irrespectively} of the initial condition \(\psi_n(0)\). Within this region, it is shown in the SM, that the elements $v_{1,n}\equiv \bra{n}\ket{v_1}$ are given by
\begin{equation}
    \label{one_defect_eigenmode}
    v_{1,n} \propto \sin\left(\frac{\pi}{2L+1}|n - q|\right),
\end{equation}
in excellent agreement with the numerical calculations, for large enough $V$.
\begin{figure}{}
    \centering
    \includegraphics[width=1\columnwidth]{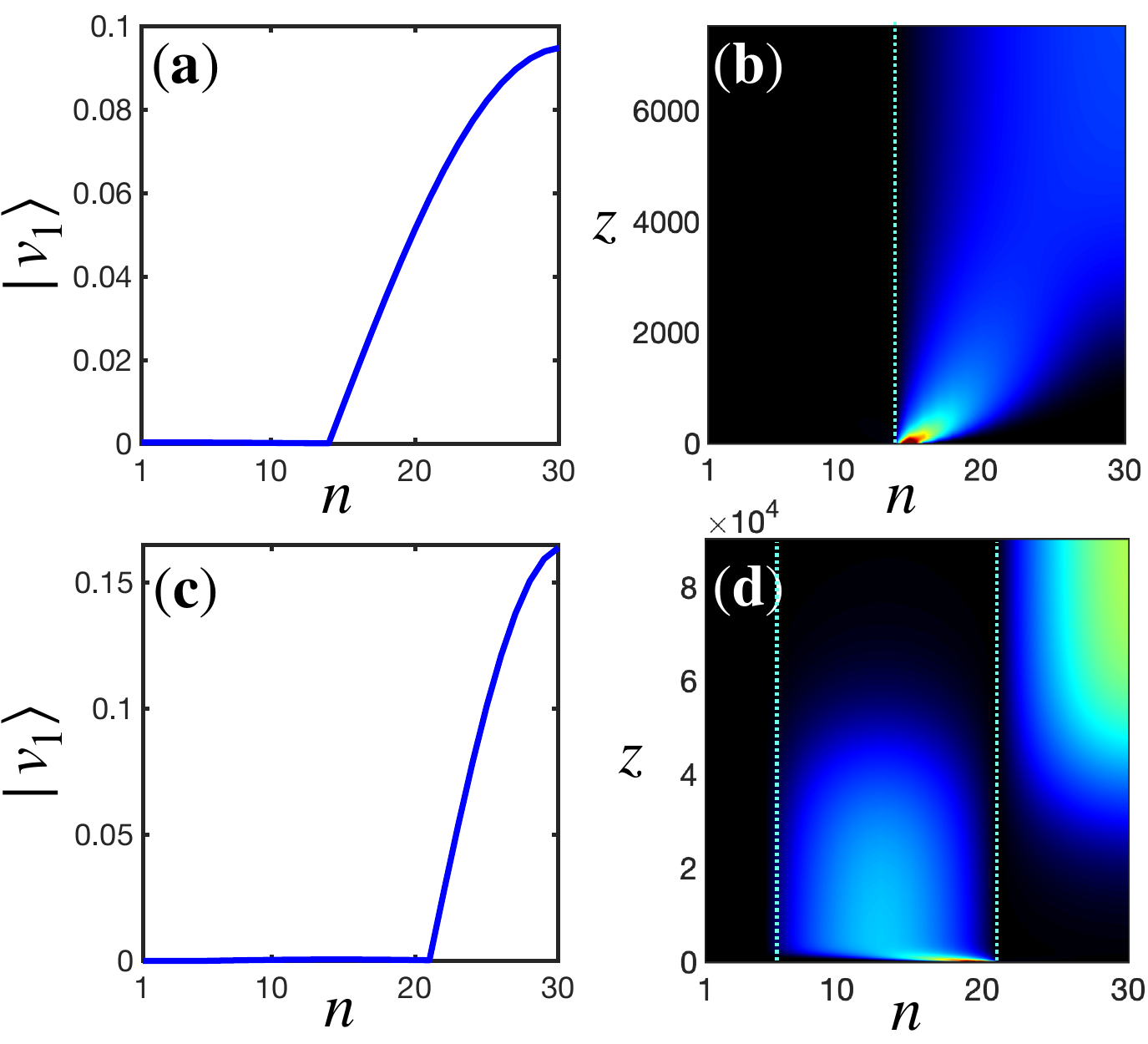}
    \caption{For a lattice of \(N = 30\) sites with dephasing period \(l = 0.01\). (a)–(b): Single defect \(\epsilon_n = 0.5\,i\,\delta_{n,14}\); (c)–(d): Two defects \(\epsilon_n = 0.5\,i\,\delta_{n,5} + 0.5\,i\,\delta_{n,21}\). (a)/(c): Amplitudes \(|v_{1,n}|\); (b)/(d): Evolution of the averaged probability density \(P(z)\), normalized at each \(z\), for initial conditions \(\psi_{n}(0) = \delta_{n,15}\) and \(\psi_{n}(0) = \delta_{n,20}\), respectively. Defect sites are indicated by blue dashed lines in (b)/(d).}
    \label{fig_1}
\end{figure}{}

A pertinent example is shown in Fig.~2, for a lattice of \(N = 30\) sites with an imaginary defect \(\epsilon_n = 0.5\,i\,\delta_{n,14}\) and \(l = 0.01\). The lossy defect induces a localized eigenmode with eigenvalue \(\beta_{\text{def}} = \beta_{30} = 0.9898 \), distinctly outside the unperturbed band (ranging between \(0.9996\)–\(1\) ). As predicted analytically, the eigenmodes \(\ket{v_j}\) exhibit spatial asymmetry, while the eigenmode \(\ket{v_1}\) corresponding to the largest eigenvalue aligns perfectly with Eq.~(\ref{one_defect_eigenmode}) [Fig.~2(a)] \cite{SM}. Consequently, despite symmetric couplings, the averaged probability density exhibits asymmetric propagation toward \(\ket{v_1}\), as illustrated in Fig.~2(b) for the single-channel initial excitation \(\psi_{n}(0) = \delta_{n,15}\). This dephasing-induced asymmetry in the dynamics forms the basis for engineering dynamical {incoherent} skin effects in non-Hermitian systems.

\textit{Two or more defects and dynamical skin effects.}---
We now consider a lattice with two identical lossy defects at positions \(q_1, q_2\), i.e., \(\epsilon_n = iV(\delta_{n,q_1} + \delta_{n,q_2})\) with \(V > 0\). As before, the defect-induced eigenvalues follow Eq.~(\ref{defect_eigenvalue}). The eigenstate \(\ket{v_1}\) associated with the largest eigenvalue \(\beta_1\) depends strongly on the defect positions. Physically, the lattice partitions into three segments: the inter-defect region of length \(M = q_2 - q_1 - 1\), and the left/right edge segments \(E_L = q_1 - 1\), \(E_R = N - q_2\). As shown analytically under the hard-wall approximation \cite{SM}, asymptotic localization occurs at the larger edge if \(M < 2\max\{E_L, E_R\}\) (forming an \emph{edge-well} state), and within the inter-defect region if \(M > 2\max\{E_L, E_R\}\) (forming a \emph{bulk-well} state). Quantitatively, the eigenmode's nonzero region is given by:
\begin{equation}
\resizebox{\columnwidth}{!}{$
\label{triplet}
v_{1,n} \propto
\begin{cases}
\sin\!\left[\dfrac{\pi\,(n - q_2)}{2(N - q_2)+1}\right], & M < 2E_R \text{ with } E_R = \max\{E_L,\,E_R\}, \\[1em]
\sin\!\left[\dfrac{\pi\,(q_1-n)}{2(q_1 - 1)+1}\right], & M < 2E_L \text{ with } E_L = \max\{E_L,\,E_R\}, \\[1em]
\sin\!\left[\dfrac{\pi\,(n - q_1)}{q_2 - q_1}\right], & M > 2\max\{E_L, E_R\}.
\end{cases}$}
\end{equation}

A pertinent example of this behavior is shown in the bottom row of Fig.~2, which depicts the evolution of the probability density \(P(z)\) for an initial excitation \(\psi_n(0) = \delta_{n,20}\) on a lattice of \(N = 30\) sites with two identical lossy defects (\(V = 0.5\)) and a dephasing period \(l = 0.01\). For defects placed at \(q_1 = 5\) and \(q_2 = 21\), we have \(M = 15\), \(E_L = 4\), and \(E_R = 9\), satisfying \(M < 2E_R\); thus, the eigenmode \(\ket{v_1}\) localizes at the right edge [Fig.~2(c)], and the asymptotic probability distribution \(P_f\) is confined near the right boundary [Fig.~2(d)]. Similar examples for left-edge localization are provided in \cite{SM}. Fig.~2(d) shows a jump in the probability distribution, marking the crossover from an initially dominant eigenmode to the asymptotically dominant state \(\ket{v_1}\). Notably, when the defect's separation is sufficiently large, satisfying \(M > 2E_L\), the wavefunction localizes between the two defects, forming a bulk-well state with defects effectively acting as impenetrable walls \cite{SM}.

Generalizing the two-defect case, we now consider an \(N\)-site lattice with \(r\) lossy defects of strength \(V\), placed at sites \(q_1, \dots, q_r\). We define the edge-well lengths as \(E_L = q_1 - 1\) and \(E_R = N - q_r\), and the inter-defect bulk-well lengths as \(M_k = q_{k+1} - q_k - 1\), \(k=1,\dots,r-1\). Denoting \(M_{\max} \equiv \max_{1\le k\le r-1} M_k = q_{j+1} - q_j - 1\), the eigenstate \(\ket{v_1}\) is given by
\begin{equation}
\resizebox{\columnwidth}{!}{$
v_{1,n}  \propto
\begin{cases}
\sin\!\left[\dfrac{\pi\,(n-q_r)}{2(N-q_r)+1}\right], & M_{\max} < 2E_R \text{ with } E_R = \max\{E_L,\,E_R\}, \\[1em]
\sin\!\left[\dfrac{\pi\,(q_1-n)}{2(q_1-1)+1}\right], & M_{\max} < 2E_L \text{ with } E_L = \max\{E_L,\,E_R\}, \\[1em]
\sin\!\left[\dfrac{\pi\,(n-q_j)}{q_{j+1}-q_j}\right], & M_{\max} > 2\max\{E_L,\,E_R\},
\end{cases}
$}
\end{equation}
in the respective edge or bulk segment where the mode is nonzero. This expression illustrates clearly how the relative positions of multiple defects determine whether the asymptotic dominant mode manifests as an edge- or bulk-well state.

A pertinent engineered configuration is presented in the top row of Fig.~3 for an \(N=30\)-site lattice with four identical lossy defects (\(V=0.5\)) located at sites \(q_i = \{3,10,18,26\}\) and dephasing period \(l=0.01\). Here, since \(M_{\text{max}}<2E_{R}\), the eigenmode \(\ket{v_1}\) forms a tightly confined right edge-state [Fig.~3(a)]. In Fig.~3(b) we show the evolution of the average probability density \(P(z)\) for an initial excitation \(\psi_{n}(0)=\delta_{n,1}\). Notably, the probability density shifts toward the right edge through a sequence of directional jumps, demonstrating a pronounced dynamical {incoherent} skin effect. This effect in the asymptotic probability density \(P_f\) arises irrespective of the initial excitation and results from the synergistic action of dephasing and the non-Hermitian lossy defects. Without dephasing, the asymptotic dynamics are dominated by the Lifshitz tails of the spectrum~\cite{Longhi2025}, which correspond to extended states, yielding diffraction across the lattice (see \cite{SM}). Thus, this defect-engineered skin effect emerges exclusively under incoherent dynamics. {Notably, since the coexistence of real disorder leaves $\mathcal{S}$ remains unchanged up to $\mathcal{O}(l^2)$, the reported incoherent skin effect is robust to any level of disorder.} The antagonism of this phenomenon with the conventional non-Hermitian skin effect in HN lattices under dephasing is studied next. 
\begin{figure}
    \centering
    \includegraphics[width=1\columnwidth]{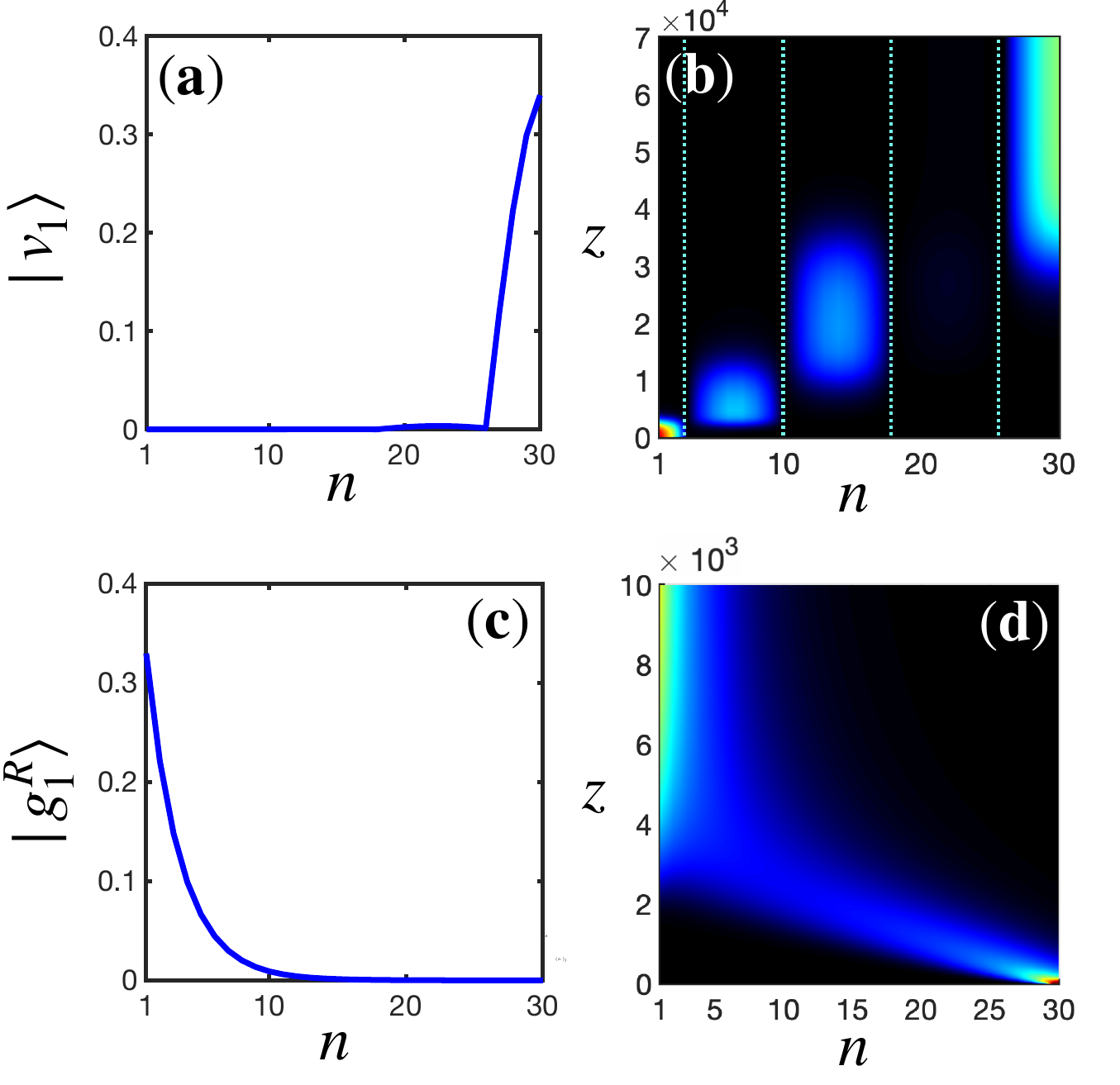}
    \caption{Dynamical skin effects under incoherent dynamics in a lattice of $N=30$ sites with dephasing period $l=0.01$, top row: symmetric couplings, defects with $V=0.5$ applied at sites $q_j=\{3,10,18,26\}$, bottom row: asymmetric couplings with $h=0.2$, no defects; (a)/(c) Amplitude $\lvert v_{1,n}\rvert$/$\lvert g_{1,n}^{R}\rvert$; (b)/(d) Evolution of the averaged probability density $P(z)$, for initial condition $\psi_{n}(0)=\delta_{n,1}$ for (b) and $\psi_{n}(0)=\delta_{n,30}$ for (d), normalized at each $z$. In (b) the lattice sites corresponding to defects are marked with blue dashed lines.}

    \label{fig_1}
\end{figure}

\textit{Dephasing in Hatano-Nelson lattices}---
\begin{figure*}
    \centering
    \includegraphics[width=1\textwidth]{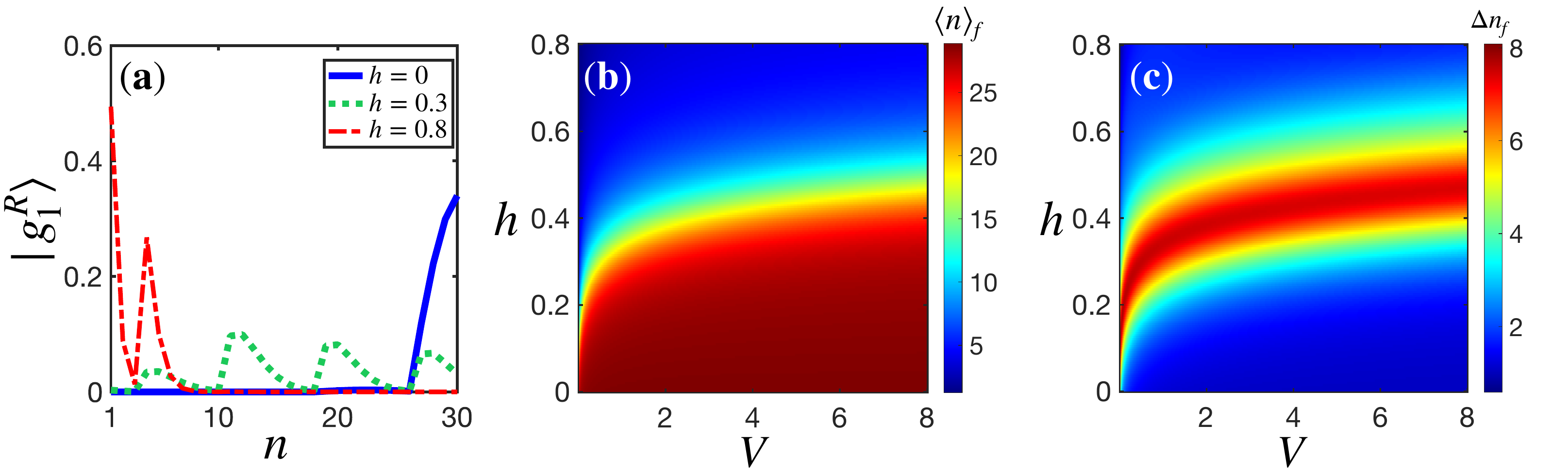}
    \caption{Interplay between non-Hermitian skin effects in a lattice of $N=30$ sites with defects of strength $V=0.5$ placed at sites $q_j=\{3,10,18,26\}$ under dephasing $l=0.01$; (a) amplitude $\lvert g_{1,n}^{R}\rvert$ for $h=0$ (blue), $h=0.3$ (green) and $h=0.8$ (red); (b) asymptotic mean position $\langle n\rangle_f$; (c) position uncertainty $\Delta n_f$ as functions of $h$ and $V$.}

    \label{fig_1}
\end{figure*}
Here, we consider a HN lattice with OBCs, under dephasing, schematically shown in Fig. 1(b). Its Hamiltonian \(\tilde H\) reads
\begin{equation} 
\label{hatano}
\tilde H = \sum_{n=1}^{N} \epsilon_{n} \ket{n}\bra{n} + \sum_{n=1}^{N-1}\bigl(e^{-h}\ket{n+1}\bra{n} + e^{h}\ket{n}\bra{n+1}\bigr),
\end{equation}
where \(h>0\) is   introducing non-Hermiticity via hopping asymmetry.
Since \(\tilde H\) is non‐symmetric, the incoherent propagator \(\Sigma\) is non‐Hermitian. Its right‐ and left‐eigenvalue problems read
as $\Sigma\ket{g_{j}^{R}} = \lambda_{j}\ket{g_{j}^{R}} $ and $\Sigma^{\dagger}\ket{g_{j}^{L}} = \lambda_{j}^{*}\ket{g_{j}^{L}}$ respectively
where the biorthogonality condition \(\bra{g_{k}^{L}}g_{j}^{R}\rangle = \delta_{kj}\) holds. From \eqref{prop} it follows that
\begin{equation}
\label{NH_expansion}
\ket{P(z = a l)} = \sum_{j=1}^{N} c_{j,0}\,\lambda_{j}^{a}\,\ket{g_{j}^{R}},
\end{equation}
where \(c_{j,0}\equiv\bra{g_{j}^{L}}P(z=0)\rangle\).

The HN Hamiltonian \(\tilde H\) relates to Hamiltonian \(H\) [Eq.~\eqref{hamiltonian}] through the similarity transformation \(\tilde H=\Omega^{-1}H\Omega\), with
$
\Omega=\sum_{n=1}^{N}e^{hn}\ket{n}\bra{n}$.
It follows that the incoherent propagator \(\Sigma\) can be written as
$\Sigma_{kj}=\bigl|(e^{i\tilde H l})_{kj}\bigr|^2
=e^{2h(j-k)}\,\mathcal S_{kj}$. Hence \(\Sigma\) and \(\mathcal S\) are similar,
\(\Sigma=\Omega^{-2}\mathcal S\,\Omega^{2}\),
and thus share the same eigenvalue spectrum. 
In the absence of on-site deviations (i.e.\ \(\epsilon_n=0\) for all \(n\)), the eigenvalues \(\lambda_j\) of \(\Sigma\) coincide with the eigenvalues \(\beta_j\) of \(\mathcal S\) given in  Eq. \eqref{eigenvalues}, while the right‐ and left‐eigenvectors of \(\Sigma\),
relate to the eigenvectors \(\ket{v_j}\) as
\begin{equation}
\label{hatano_vect}
\begin{aligned}
\ket{g_j^R}&=\Omega^{-2}\ket{v_j}
\propto\sum_{n=1}^{N}e^{-2hn}\cos\!\left[\frac{(j-1)\pi}{N}\,\left(n-\tfrac12\right)\right]\ket{n},\\
\ket{g_j^L}&=\Omega^{2}\ket{v_j}
\propto\sum_{n=1}^{N}e^{2hn}\cos\!\left[\frac{(j-1)\pi}{N}\,\left(n-\tfrac12\right)\right]\ket{n}
\end{aligned}
\end{equation}
Since \(\mathcal S\) has \(\beta_1=1\) and also all other eigenvalues satisfy \(|\beta_j|\le1\), the same holds for \(\lambda_j\). It follows from \eqref{NH_expansion} that projections onto \(\{\ket{g_m^R}\}_{m\neq1}\) vanish as \(z\to\infty\). In this limit, \emph{any initial excitation} evolves into an asymmetric distribution,
$P_f=\ket{g_1^R}=\Omega^{-2}\ket{v_1}\propto\bigl(e^{-2h},e^{-4h},\dots,e^{-2hN}\bigr)^T$,
which manifests the non‐Hermitian skin effect in the incoherent regime.

In Figure~3(c) we depict the right eigenmode \(\ket{g_1^R}\) of \(\Sigma\) for a lattice with \(N=30\), \(l=0.01\), and \(h=0.2\), in agreement with the aforementioned formula. Figure~3(d) depicts the dynamics of the average probability density \(P(z)\) for the same lattice with initial excitation \(\psi_n(0)=\delta_{n,30}\). The system evolves into the asymmetric edge‐localized distribution \(P_f\propto(e^{-2h},e^{-4h},\dots,e^{-2hN})^T\).

\textit{Antagonism of Non-Hermitian Skin Effects.}---
We now consider lattices with both defects and coupling asymmetry, leading to competition between the two types of non-Hermitian skin effects identified previously in the incoherent regime.

In Fig. 4(a), we plot the asymptotic probability density \(P_f=\ket{g_{1}^{R}}\) for a lattice with \(N=30\) sites under dephasing (\(l=0.01\)) with four identical defects at \(q_i=\{3,10,18,26\}\), corresponding to Fig. 3(a)/(b).  In the symmetric case (\(h=0\)), edge localization arises solely due to defects. With introduced coupling asymmetry \(h=0.3\), \(P_f\) shifts away from pure edge localization, forming a \emph{jagged-shaped} pattern, broadly distributed with nodes at the defect positions. However, this asymmetry remains insufficient to shift localization entirely to the opposite edge. Further increasing asymmetry to \(h=0.8\) predominantly localizes the wavefunction on the left side, while the defect at site \(q_1=3\) forces the existence of a node.

To quantify the interplay between the defect-induced and coupling-induced skin effects, we calculate the averaged asymptotic mean position  \(\langle n \rangle_f\) and position uncertainty \(\Delta n_f\) \cite{unc} of the wavefunction as functions of the coupling-asymmetry parameter \(h\) and the defect strength \(V\) [Fig. 4(b)-(c)], revealing the transition between the two skin effects. As shown in Fig. 4(b), for a fixed \(V\) there exists a range of \(h\) values where the defect-induced {incoherent} skin effect remains dominant, with \(\langle n \rangle_f \sim N\). Beyond a certain threshold of \(h\), the mean position shifts and the long-term probability density becomes more extended, since \(\Delta n_f\) is increasing. A further increase in \(h\) ultimately leads to the manifestation of the coupling-induced skin effect, with \(\langle n \rangle_f\to 1\) and decrease of \(\Delta n_f\). Moreover, as evident in  Fig. 4(a)-(b), the critical value of $h$ needed for this transition, rises as the strength $V$ of defects is increased. 
Although our quantitative analysis is based on the configuration of Fig. 4(a), these conclusions hold broadly; defect-driven dynamical skin effects can be precisely engineered in incoherent lattices, with preferential localization persisting robustly until coupling asymmetry surpasses a critical threshold.

\textit{Conclusions.}---
In this work, we have identified a novel route to realizing a dynamical {incoherent} non-Hermitian skin effect without relying on asymmetric couplings. Strategically engineered configurations of lossy defects, combined with fast dephasing, give rise to asymmetric propagation via directional jumps toward the lattice edge. We also examine the incoherent dynamics of a Hatano–Nelson lattice and investigate the interplay between these two distinct skin-effect mechanisms, revealing the emergence of jagged-shaped modes. Our results open new avenues for the experimental observation of dynamical skin effects and for the theoretical exploration of dephasing as a tool for wave manipulation.


\begin{thebibliography}\
\bibitem{ElGanainy2018} R. El-Ganainy, K. G. Makris, M. Khajavikhan, Z. H. Musslimani, S. Rotter and D. N. Christodoulides, Non-Hermitian physics and PT symmetry, {Nat. Physics} \textbf{14}, 11–19 (2018).
\bibitem{Ruter2010} C. E. Rüter, K. G. Makris, R. El-Ganainy, D. N. Christodoulides, M. Segev and D. Kip, Observation of parity–time symmetry in optics, {Nat. Physics} \textbf{6}, 192 (2010).
\bibitem{Makris2008} K. G. Makris, R. El-Ganainy, D. N. Christodoulides and Z. H. Musslimani, Beam Dynamics in \(\mathcal{PT}\) Symmetric Optical Lattices, {Phys. Rev. Lett.} \textbf{100}, 103904 (2008).
\bibitem{ElGanainy2007} R. El-Ganainy, K. G. Makris, D. N. Christodoulides, and Z. H. Musslimani, Theory of coupled optical \(\mathcal{PT}\)-symmetric structures, {Opt. Lett.} \textbf{32}, 2632 (2007).
\bibitem{Musslimani2008} Z. H. Musslimani, K. G. Makris, R. El-Ganainy and D. N. Christodoulides, Optical Solitons in \(\mathcal{PT}\) Periodic Potentials, {Phys. Rev. Lett.} \textbf{100}, 030402 (2008).
\bibitem{Guo2009} A. Guo, G. J. Salamo, D. Duchesne, R. Morandotti, M. Volatier-Ravat, V. Aimez, G. A. Siviloglou, and D. N. Christodoulides, Observation of PT–Symmetry Breaking in Complex Optical Potentials, {Phys. Rev. Lett.} \textbf{103}, 093902 (2009).
\bibitem{Regensburger2012} A. Regensburger, C. Bersch, M. A. Miri, G. Onishchukov, D. N. Christodoulides and U. Peschel, Parity–time synthetic photonic lattices, {Nature (London)} \textbf{488}, 167 (2012).
\bibitem{Hodaei2014} H. Hodaei, M. A. Miri, M. Heinrich, D. N. Christodoulides and M. Khajavikhan, Parity–time–symmetric microring lasers, {Science} \textbf{346}, 975 (2014).
\bibitem{Konotop2016} V. V. Konotop, J. Yang, and D. A. Zezyulin, Nonlinear waves in \(\mathcal{PT}\)-symmetric systems, {Rev. Mod. Phys.} \textbf{88}, 035002 (2016).
\bibitem{Feng2017} L. Feng, R. El-Ganainy, and L. Ge, Non-Hermitian photonics based on parity–time symmetry, {Nat. Photon.} \textbf{11}, 752 (2017).
\bibitem{Ozdemir2019} S. K. Özdemir, S. Rotter, F. Nori, and L. Yang, Parity–time symmetry and exceptional points in photonics, {Nat. Mater.} \textbf{18}, 783 (2019).

\bibitem{Hatano1996}N. Hatano and D. R. Nelson, Localization Transitions in Non-Hermitian Quantum Mechanics, {Phys. Rev. Lett.} \textbf{77}, 570 (1996).
\bibitem{Hatano1997}N. Hatano and D. R. Nelson, Vortex pinning and non-Hermitian quantum mechanics, {Phys. Rev. B} \textbf{56}, 8651 (1997).
\bibitem{Zhang2022review}
X. Zhang, T. Zhang, M. H. Lu, and Y. F. Chen, A review on non-Hermitian skin effect, \emph{Adv. Phys.: X} \textbf{7}, 1 (2022).
\bibitem{Lee2016}
T. E. Lee, Anomalous Edge State in a Non-Hermitian Lattice, \emph{Phys. Rev. Lett.} \textbf{116}, 133903 (2016).
\bibitem{MartinezAlvarez2018}
V. M. Martinez Alvarez, J. E. Barrios Vargas, and L. E. F. Foa Torres, Non-Hermitian robust edge states in one-dimension: anomalous localization and eigenspace condensation at exceptional points, \emph{Phys. Rev. B} 
\textbf{97}, 121401(R) (2018).
\bibitem{Gong2018}
Z. Gong, Y. Ashida, K. Kawabata, K. Takasan, S. Higashikawa, and M. Ueda, topological phases of non-Hermitian systems, \emph{Phys. Rev. X} \textbf{8}, 031079 (2018).
\bibitem{Yao2018}
S. Yao and Z. Wang, Edge States and Topological Invariants of Non-Hermitian Systems, \emph{Phys. Rev. Lett.} \textbf{121}, 086803 (2018).
\bibitem{Longhi2015b}
S. Longhi, D. Gatti, and G. Della Valle, Non-Hermitian transparency and one-way transport in low-dimensional lattices by an imaginary gauge field, \emph{Phys.\ Rev.\ B} \textbf{92}, 094204 (2015).
\bibitem{Longhi2015}
S. Longhi, D. Gatti, and G. D. Valle, Robust light transport in non-Hermitian photonic lattices, \emph{Sci.\ Rep.} \textbf{5}, 13376 (2015).
\bibitem{Longhi2016}
S. Longhi, Tight-binding lattices with an oscillating imaginary gauge field, \emph{Phys.\ Rev.\ A} \textbf{94}, 022102 (2016).
\bibitem{Longhi2017a}
S. Longhi, Non-Hermitian bidirectional robust transport, \emph{Phys.\ Rev.\ B} \textbf{95}, 014201 (2017).
\bibitem{Longhi2017b}
S. Longhi, Nonadiabatic robust excitation transfer assisted by an imaginary gauge field, \emph{Phys.\ Rev.\ A} \textbf{95}, 062122 (2017).
\bibitem{Longhi2022}
S. Longhi, Non-Hermitian skin effect and self-acceleration, \emph{Phys.\ Rev.\ B} \textbf{105}, 245143 (2022).
\bibitem{Li2022}
H. Li and S. Wan, Dynamic skin effects in non-Hermitian systems, \emph{Phys.\ Rev.\ B} \textbf{106}, L241112 (2022).
\bibitem{Komis2023}
I. Komis, Z. H. Musslimani, and K. G. Makris, Skin solitons, \emph{Opt.\ Lett.} \textbf{48}, 6525–6528 (2023).
\bibitem{Kokkinakis2024}
E. T. Kokkinakis, K. G. Makris, and E. N. Economou, Anderson localization versus hopping asymmetry in a disordered lattice, \emph{Phys.\ Rev.\ A} \textbf{110}, 053517 (2024).

\bibitem{Yi2020}
Y. Yi and Z. Yang, Non-Hermitian skin modes induced by on-site dissipations and chiral tunneling effect, \emph{Phys.\ Rev.\ Lett.} \textbf{125}, 186802 (2020).
\bibitem{Li2020}
L. Li, C. H. Lee, S. Mu, and J. Gong, Critical non-Hermitian skin effect, \emph{Nat.\ Commun.} \textbf{11}, 5491 (2020).
\bibitem{Guo2021}
C.-X. Guo, C.-H. Liu, X.-M. Zhao, Y. Liu, and S. Chen, Exact solution of non-Hermitian systems with generalized boundary conditions: size-dependent boundary effect and fragility of skin effect, \emph{Phys.\ Rev.\ Lett.} \textbf{127}, 116801 (2021).
\bibitem{Jiang2023}
C. Jiang, Y. Liu, X. Li, Y. Song, and S. Ke, Twist-induced non-Hermitian skin effect in optical waveguide arrays, \emph{Appl.\ Phys.\ Lett.} \textbf{123}, 151101 (2023).
\bibitem{Huang2024}
X. Huang, Y. Li, G.-F. Zhang, and Y.-C. Liu, Non-Hermitian skin effect and nonreciprocity induced by dissipative couplings, \emph{Phys.\ Rev.\ A} \textbf{109}, L021503 (2024).
\bibitem{Lei2024}
Z. Lei, C. H. Lee, and L. Li, Activating non-Hermitian skin modes by parity–time symmetry breaking, \emph{Commun.\ Phys.} \textbf{7}, 100 (2024). 
\bibitem{Samanta2025}R. Samanta, A. Chakrabarty, and S. Datta, Engineering unique localization transition with coupled Hatano–Nelson chains, SciPost Phys. Core \textbf{8}, 033 (2025).
\bibitem{Ganguly2025}S. Ganguly and S. K. Maiti, Persistent current in a non-Hermitian Hatano–Nelson ring: Disorder-induced amplification, Phys. Rev. B \textbf{111}, 195418 (2025).
\bibitem{Ammari2025}H. Ammari, S. Barandun, J. Cao, B. Davies, E. O. Hiltunen, and P. Liu, The non-Hermitian skin effect with three-dimensional long-range coupling, J. Eur. Math. Soc. (2025), doi:10.4171/JEMS/1685.
\bibitem{Molignini2023}P. Molignini, O. Arandes, and E. J. Bergholtz, Anomalous skin effects in disordered systems with a single non-Hermitian impurity, Phys. Rev. Res. \textbf{5}, 033058 (2023).
\bibitem{Yokomizo2021} K. Yokomizo and S. Murakami, Scaling rule for the critical non-Hermitian skin effect, {Phys. Rev. B} \textbf{104}, 165117 (2021).\
\bibitem{Liu2021} Y. Liu, Y. Zeng, L. Li, and S. Chen, Exact solution of the single impurity problem in nonreciprocal lattices: Impurity-induced size-dependent non-Hermitian skin effect, {Phys. Rev. B} \textbf{104}, 085401 (2021).
\bibitem{FuZhang2023} Y. Fu and Y. Zhang, Hybrid scale-free skin effect in non-Hermitian systems: A transfer matrix approach, {Phys. Rev. B} \textbf{108}, 205423 (2023).

\bibitem{song2019}
F.~Song, S.~Yao, and Z.~Wang, Non-Hermitian skin effect and chiral damping in open quantum systems, \emph{Phys. Rev. Lett.} \textbf{123}, 170401 (2019).

\bibitem{kim2024}
B.~H.~Kim, J.-H.~Han, and M.~J.~Park, Collective non-Hermitian skin effect: point-gap topology and the doublon-holon excitations in non-reciprocal many-body systems, \emph{Commun. Phys.} \textbf{7}, 73 (2024).

\bibitem{Mattiello2025}V. M. Mattiello, V. L. Quito, and E. Miranda, Asymptotically exact solution of the non-Hermitian disordered interacting Hatano–Nelson chain, arXiv:2509.16309 [cond-mat.str-el] (2025).

\bibitem{Suthar2025}K. Suthar, Boundary-driven many-body phase transitions in a non-Hermitian disordered fermionic chain, Phys. Rev. B \textbf{111}, 064202 (2025).

\bibitem{Lee2019b}
C. H. Lee, L. Li, and J. Gong, Hybrid Higher-Order Skin-Topological Modes in Nonreciprocal Systems, \emph{Phys.\ Rev.\ Lett.} \textbf{123}, 016805 (2019).
\bibitem{Kawabata2020}
K. Kawabata, M. Sato, and K. Shiozaki, Higher-order non-Hermitian skin effect, \emph{Phys.\ Rev.\ B} \textbf{102}, 205118 (2020).
\bibitem{zhang2021}
X. Zhang, Y. Tian, J.-H. Jiang, M.-H. Lu, and Y.-F. Chen, Observation of higher-order non-Hermitian skin effect, \emph{Nat.\ Commun.} \textbf{12}, 5377 (2021).
\bibitem{Zhang2022}
K. Zhang, Z. Yang, and C. Fang, Universal non-Hermitian skin effect in two and higher dimensions, \emph{Nat.\ Commun.} \textbf{13}, 2496 (2022).
\bibitem{Kokkinakis2025}
E. T. Kokkinakis, I. Komis, and K. G. Makris, Self-trapping and skin solitons in two-dimensional non-Hermitian lattices, \emph{arXiv:2505.12463} (2025).

\bibitem{Li2023}
Y. Li, C. Lu, S. Zhang, and Y.-C. Liu, Loss-induced Floquet non-Hermitian skin effect, \emph{Phys.\ Rev.\ B} \textbf{108}, L220301 (2023).
\bibitem{Ke2023}
S. Ke, W. Wen, D. Zhao, and Y. Wang, Floquet engineering of the non-Hermitian skin effect in photonic waveguide arrays, \emph{Phys.\ Rev.\ A} \textbf{107}, 053508 (2023). 
\bibitem{Apostolidis2025}A. Apostolidis, N. S. Nye, N. V. Kantartzis, D. N. Christodoulides, and G. G. Pyrialakos, Kin-effect localization and maximal-order exceptional points in reciprocal Floquet lattices, Commun. Phys. \textbf{8}, 358 (2025).
\bibitem{yao2018}
S.~Yao and Z.~Wang, Edge states and topological invariants of non-Hermitian systems, \emph{Phys. Rev. Lett.} \textbf{121}, 086803 (2018).
\bibitem{Lee2019}
C. H. Lee and R. Thomale, Anatomy of skin modes and topology in non-Hermitian systems, \emph{Phys.\ Rev.\ B} \textbf{99}, 201103(R) (2019).
\bibitem{okuma2020}
N.~Okuma, K.~Kawabata, K.~Shiozaki, and M.~Sato, Topological origin of non-Hermitian skin effects, \emph{Phys. Rev. Lett.} \textbf{124}, 086801 (2020).


\bibitem{Weidemann2020}S. Weidemann, M. Kremer, T. Helbig, T. Hofmann, A. Stegmaier, M. Greiter, R. Thomale and A. Szameit, Topological funneling of light, {Science} \textbf{368}, 311 (2020).
\bibitem{Gao2023}
Z. Gao, X. Qiao, M. Pan, S. Wu, J. Yim, K. Chen, B. Midya, L. Ge, and L. Feng, Two-Dimensional Reconfigurable Non-Hermitian Gauged Laser Array, \emph{Phys.\ Rev.\ Lett.} \textbf{130}, 263801 (2023).
\bibitem{Liu2022}
Y. G. Liu, Y. Wei, O. Hemmatyar, G. G. Pyrialakos, P. S. Jung, D. N. Christodoulides, and M. Khajavikhan, Complex skin modes in non-Hermitian coupled laser arrays, \emph{Light: Sci.\ Appl.} \textbf{11}, 336 (2022). 

\bibitem{Zhang2021a}
X. Zhang, Y. Tian, J. H. Jiang, M. H. Lu, and Y. F. Chen, Observation of higher-order non-Hermitian skin effect, \emph{Nat.\ Commun.} \textbf{12}, 1–8 (2021). 
\bibitem{Zhang2021b}
L. Zhang, Y. Yang, Y. Ge, Y.-J. Guan, Q. Chen, Q. Yan, F. Chen, R. Xi, Y. Li, D. Jia, S.-Q. Yuan, H.-X. Sun, H. Chen, and B. Zhang, Acoustic non-Hermitian skin effect from twisted winding topology, \emph{Nat.\ Commun.} \textbf{12}, 1–7 (2021). 


\bibitem{li2024}
Z.~Li, L.-W.~Wang, X.~Wang, Z.-K.~Lin, G.~Ma, and J.-H.~Jiang, Observation of dynamic non-Hermitian skin effects, \emph{Nat. Commun.} \textbf{15}, 50776 (2024).

\bibitem{Zhao2025}
E. Zhao, Z. Wang, C. He, T. Fung, J. Poon, K. K. Pak, Y.-J. Liu, P. Ren, X.-J. Liu, and G.-B. Jo, Two-dimensional non-Hermitian skin effect in an ultracold Fermi gas, 


\bibitem{logan_1987}
D. E. Logan and P. G. Wolynes, Dephasing and Anderson localization in topologically disordered systems, \emph{Phys.\ Rev.\ B} \textbf{36}, 4135 (1987).
\bibitem{evensky_1990}
D. A. Evensky, R. T. Scalettar, and P. G. Wolynes, Localization and dephasing effects in a time-dependent Anderson Hamiltonian, \emph{J.\ Phys.\ Chem.} \textbf{94(3)}, 1149 (1990).
\bibitem{flores_1999}
J. C. Flores, Diffusion in disordered systems under iterative measurement, \emph{Phys.\ Rev.\ B} \textbf{60}, 30 (1999).
\bibitem{yamada_1999}
H. Yamada and K. S. Ikeda, Dynamical delocalization in one-dimensional disordered systems with oscillatory perturbation, \emph{Phys.\ Rev.\ E} \textbf{59}, 5214 (1999).
\bibitem{gurvitz_2000}
S. A. Gurvitz, Delocalization in the Anderson Model due to a Local Measurement, \emph{Phys.\ Rev.\ Lett.} \textbf{85}, 812 (2000).
\bibitem{kosik_2006}
J. Košík, V. Bužek, and M. Hillery, Quantum walks with random phase shifts, \emph{Phys.\ Rev.\ A} \textbf{74}, 022310 (2006).
\bibitem{schreiber_2011}
A. Schreiber, K. N. Cassemiro, V. Potoček, A. Gábris, I. Jex, and C. Silberhorn, Decoherence and Disorder in Quantum Walks: From Ballistic Spread to Localization, \emph{Phys.\ Rev.\ Lett.} \textbf{106}, 180403 (2011).
\bibitem{moix_2013}
J. M. Moix, M. Khasin, and J. Cao, Coherent quantum transport in disordered systems: I. The influence of dephasing on the transport properties and absorption spectra on one-dimensional systems, \emph{New J.\ Phys.} \textbf{15}, 085010 (2013).
\bibitem{znidaric_2013}
M. Žnidarič and M. Horvat, Transport in a disordered tight-binding chain with dephasing, \emph{Eur.\ Phys.\ J.\ B} \textbf{86}, 67 (2013).
\bibitem{kamiya_2015}
N. Kamiya, Quantum-to-classical reduction of quantum master equations, \emph{Prog.\ Theor.\ Exp.\ Phys.} \textbf{2015}, 043A02 (2015).
\bibitem{krapivsky_2014}
P. L. Krapivsky, J. M. Luck, and K. Mallick, Survival of classical and quantum particles in the presence of traps, \emph{J. Stat.\ Phys.} \textbf{154}, 1430 (2014).
\bibitem{longhi_2024_prl}S. Longhi, Dephasing-Induced Mobility Edges in Quasicrystals, {Phys. Rev. Lett.} \textbf{132}, 236301 (2024).
\bibitem{longhi_2024_ol}S. Longhi, Robust Anderson transition in non-Hermitian photonic quasicrystals, {Opt. Lett.} \textbf{49}, 1373 (2024).
\bibitem{Kokkinakis2024b}
E.~T.~Kokkinakis, K.~G.~Makris, and E.~N.~Economou, Dephasing-induced jumps in non-Hermitian disordered lattices, {Phys. Rev. B} \textbf{111}, 214204 (2025). 
\bibitem{Tzortzakakis2021} A. F. Tzortzakakis, K. G. Makris, A. Szameit, and E. N. Economou, Transport and spectral features in non-Hermitian open systems, {\it Phys. Rev. Research} \textbf{3}, 013208 (2021).
\bibitem{Weidemann2021} S. Weidemann, M. Kremer, S. Longhi, and A. Szameit, Coexistence of dynamical delocalization and spectral localization through stochastic dissipation, {\it Nat. Photonics} \textbf{15}, 576 (2021).
\bibitem{Leventis2022} A. Leventis, K. G. Makris, and E. N. Economou, Non-Hermitian jumps in disordered lattices, {\it Phys. Rev. B} \textbf{106}, 064205 (2022).

\bibitem{SM} See Supplementary Material for details: (I) Incoherent propagation, including (A) derivation of propagation equation; (B) eigenspectrum of propagator for lattices with real on-site deviations; (C) eigenspectrum of propagator and dynamics for a single imaginary defect; (D) eigenspectrum of propagator and dynamics for two imaginary defects; (F) dynamics for single random-phase realizations; and (II) coherent propagation, covering (A) spectral symmetries of Hamiltonians with defects and (B) corresponding long-term dynamics.
\bibitem{longhi_2024_lsa}S. Longhi, Incoherent non-Hermitian skin effect in photonic quantum walks, {Light: Sci. Appl.} \textbf{13}, 95 (2024).
\bibitem{Longhi2025} S. Longhi, Lifshitz tail states in non-Hermitian disordered photonic lattices, \textit{Opt. Lett.} \textbf{50}, 746–749 (2025).
\bibitem{unc}
We define the asymptotic mean position and position uncertainty as follows:
$	\langle n \rangle_{f} \equiv \sum_{n=1}^{N} n \cdot P_{n,f}$ and $\Delta n_{f} \equiv \sqrt{\langle n^2 \rangle_{f} - \langle n \rangle_{f}^2}$ with $\langle n^2 \rangle_{f} \equiv \sum_{n=1}^{N} n^{2} \cdot P_{n,f}$ and \(P_{n,f}\equiv \bra{n}\ket{P_f}\)
\end{thebibliography}
\end{document}